\documentclass{svproc}
\usepackage{url}
\usepackage{cite}
\usepackage[colorlinks=true,
linkcolor=blue,
citecolor=blue,
urlcolor=blue,
filecolor=blue,
pdfpagemode=None,
pdfstartview=FitH]{hyperref}

\usepackage{graphicx}
\usepackage{amsmath,amsfonts,amssymb}
\usepackage{bm}
\usepackage{subcaption}
\usepackage{siunitx}
\usepackage{tikz,pgfplots}
\usetikzlibrary{positioning, arrows.meta}
\tikzset{%
	myarrow/.style = {-Stealth, shorten >=5pt}
}
\newcommand{\mvect}[1]{\mathbf{#1}}

\newcommand{\pd}[2]{\frac{\partial #1}{\partial #2}}

\newenvironment{rcases}
{\left.\begin{aligned}}
	{\end{aligned}\right\rbrace}

\begin{document}
\mainmatter              % start of a contribution
\title{Topology optimization of pressure-loaded multi-material structures}
\titlerunning{Topology optimization of pressure-loaded multi-material structures}  % abbreviated title (for running head)
\author{Prabhat Kumar}
\authorrunning{P. Kumar} % abbreviated author list (for running head)
\institute{Department of Mechanical and Aerospace Engineering, Indian Institute of Technology Hyderabad, Telangana 502285, India,\\
\email{pkumar@mae.iith.ac.in}}

\maketitle              % typeset the title of the contribution

\begin{abstract}
 Permitting multiple materials within a topology optimization setting increases the search space of the technique, which facilitates obtaining high-performing and efficient optimized designs. Structures with multiple materials involving fluidic pressure loads find various applications. However, dealing with the design-dependent nature of the pressure loads is challenging in topology optimization that gets even more pronounced with a multi-material framework. This paper provides a density-based topology optimization method to design fluidic pressure loadbearing multi-material structures. The design domain is parameterized using hexagonal elements as they ensure nonsingular connectivity. Pressure modeling is performed using the Darcy law with a conceptualized drainage term. The flow coefficient of each element is determined using a smooth Heaviside function considering its solid and void states. The consistent nodal loads are determined using the standard finite element methods. Multiple materials is modeled using the extended SIMP scheme. Compliance minimization with volume constraints is performed to achieve optimized loadbearing structures. Few examples are presented to demonstrate the efficacy and versatility of the proposed approach. The optimized results contain the prescribed amount of different materials.
\end{abstract}

\section{Introduction}
Topology optimization (TO) is a systematic design technique that determines the optimized material distribution within a  given design domain by extremizing an objective with physical/geometrical constraints. The method has already proven its relevance and utility in various applications \cite{sigmund2013topology,van2013level}. TO can be used for problems that involve partial differential equations (PDEs) to represent their physics. One solves the associated PDE in each TO iteration; thus, a  ubiquitous option is to use finite element analysis. Herein, \texttt{HoneyMesher} code provided with \texttt{HoneyTop90} in~\cite{kumar2022honeytop90} is employed to parameterize the design domains by hexagonal elements. Such elements provide non-singular connectivity between two juxtaposed elements; thus, it offers checkerboard-free and point-connection-free optimized designs~\cite{kumar2022honeytop90,langelaar2007use,saxena2011topology,talischi2009honeycomb}. Wachspress shape functions are used to model the honeycomb tessellations used to represent the designs. The shape functions are pretty rich compared to bilinear shape functions employed for quadrilateral elements \cite{kumar2022honeytop90,wachspress1975rational,sukumar2004conforming}. Each element is associated with the design variable that decide different material phases for the element~\cite{kumar2022honeytop90}. 

Lightweight and high-performing structures are typically constructed using different materials and find applications in aircraft and the military. When designing such structures using TO, it is deemed to include a multi-material formulation within the TO setting. That is, on parameterized level, the optimizer can choose to provide different materials to an element. That generally increases the optimization search space, which facilitates a better-performing optimized structure, perhaps have lightweight. In addition, with the unprecedented advancement of additive manufacturing techniques, optimized structures with multi-material can be readily fabricated. The loads acting on such structures can be either design-independent or design-dependent (e.g., fluidic pressure load~\cite{kumar2020topology,kumar2020topology3Dpressure,kumar2023TOPress}, self-weight \cite{kumar2022topology}, etc.). In various applications, structures experience pressure loads that are challenging to handle in a TO setting \cite{Hammer2000,kumar2020topology,kumar2023TOPress}. In addition, to the best of author's knowledge, to date, only the method presented in~\cite{sivapuram2021design} optimizes design-dependent pressure loadbearing structures with many candidate materials using the integer programming method. This paper provides a TO approach to design pressure-loaded multi-material structures described using honeycomb tessellation with the well-cherished density-based method.

Sigmund and Torquato~\cite{sigmund1997design}  proposed a multi-material based TO approach for designing materials with extreme thermal expansion. They extended the Solid Isotropic Material with Penalization (SIMP) scheme for the multiple materials' modeling. The procedure requires additional design variables per the number of candidate materials. The Recursive Multiphase Material Interpolation (RMMI) with mass constraints is proposed for many candidate materials TO in~\cite{gao2011mass}. To design multi-material microstructures with different bulk moduli, Gibiansky and Sigmund~\cite{gibiansky2000multiphase} used the extended SIMP scheme, whereas Fujii et al.~\cite{fujii2001composite}  used a homogenization-based method. An approach using a peak function was presented by Yin and Ananthasuresh~\cite{yin2001topology}. The method does not require additional design variables as the number of candidate materials increases; however, a gradual adjustment of the parameters is needed to achieve multiple peaks. Zuo and Saitou~\cite{zuo2017multi} presented an ordered SIMP scheme that uses one design variable per element even for various materials; however, the interpolation scheme has non-differentiable nature. An approach using a neural network for multi-material TO was presented in~\cite{chandrasekhar2021multi}. Sivapuram et al.~\cite{sivapuram2021design} offered a TO process for multiple materials structures using integer programming and the extended SIMP approach. They also optimized multi-material pressure loadbearing structures in their paper. Cherri{\`e}re et al.~\cite{cherriere2022multi} presented a multi-material TO approach using the Wachspress interpolation. Besides the SIMP method, multi-material TO using the level set method~\cite{wang2004color,liu2020multi}, the phase field approach~\cite{tavakoli2014multimaterial} and ESO method~\cite{zhao2019topology} also exist. We employ the extended SIMP scheme for multi-material modeling for designing pressure loadbearing structures herein.

Pressure loads alter their direction, magnitude, and/or location as TO advances; thus, they are challenging to model in a TO setting \cite{kumar2020topology,Hammer2000}. Hammer and Olhoff~\cite{Hammer2000} were the first to present a TO method to design structures involving pressure loads. The technique uses iso-density contour curves/surfaces to detect the loading surface. Fuchs and Shemesh~\cite{fuchs2004} presented a method wherein they used an additional variable for pressure boundary evolution. Zhang et al.~\cite{zhang2008new} proposed an element-based approach to locate the pressure boundary. Lee and Martins~\cite{lee2012structural} presented a method that is not dependent upon the initial and final points for the pressure curves. Ibhadode et al.~\cite{ibhadode2020topology} provided an approach with boundary identification and load evolution. The method mentioned above either evaluated load sensitivities using the finite difference method or completely neglected them from overall sensitivities. However, it is demonstrated in~\cite{kumar2020topology,kumar2023TOPress} that the load sensitivities affect the final topologies. A fictitious thermal loading approach was presented by Chen and Kikuchi~\cite{Chen2001}. Sigmund and Clausen~\cite{Sigmund2007} presented an approach using the mixed finite element method that needs to satisfy the Babuska-Brezzi condition~\cite{zienkiewicz2005finite}. A pseudo electrical potential scheme was proposed by Zheng et al.~\cite{Zheng2009}. Picelli et al.~\cite{picelli2019topology} used the level-set method in their approach. Kumar et al.~\cite{kumar2020topology} presented a density-based TO approach. They solved pressure loadbearing structures and pressure-actuated compliant mechanisms and considered load sensitivities within the approach. The approach~\cite{kumar2020topology} is successfully extended for 3D problems in~\cite{kumar2020topology3Dpressure}, with the robust TO formulation in~\cite{kumar2022topological}, with the material masks overlay approach for 0-1 optimized designs in~\cite{kumar2022improved} and recently, a 100-line MATLAB code, \texttt{TOPress}, is presented in~\cite{kumar2023TOPress}. For a comprehensive list of TO approaches involving design-dependent pressure loads, one may refer to~\cite{kumar2020topology,picelli2019topology,kumar2020topology3Dpressure} and references therein. We adopt the method presented in \cite{kumar2020topology} to model the pressure loads with honeycomb tessellation~\cite{kumar2022honeytop90}.

The remainder of the manuscript is structured as follows. Sec.~\ref{Sec:MultimaterialModeling} provides the multi-material modeling formulation employed in this paper. Pressure load modeling is presented in Sec.~\ref{Sec:PressLoadModel}; for a detailed description, one can refer to~\cite{kumar2020topology}. Sec.~\ref{Sec:TOformulation} presents the topology optimization formulation for the multi-material. The success and efficacy of the proposed approach are demonstrated in Sec.~\ref{Sec:NumExDiss} by showing different numerical results and discussions. Lastly, Sec.~\ref{Sec:ConRem} draws the concluding remarks.

\section{Multi-material Modeling}\label{Sec:MultimaterialModeling}
The different methods to model multiple materials with various TO settings are  introduced in the previous section. We use a density-based TO method and the extended SIMP interpolation scheme to formulate the multiple materials. The number of design variables assigned to each element is the same as that of the candidate materials. The modified SIMP scheme for a two-phase (single material) TO is
\begin{equation}~\label{Eq:oneMaterial}
	E_i = E_\mathrm{min}+{\tilde{\rho}_i}^p (E_1 - E_\mathrm{min}) = (1 -\tilde{\rho}_i^p)E_\mathrm{min} + \tilde{\rho}_i^pE_1
\end{equation}
where $E_\text{min}$ and $E_1$ are Youngs' moduli of the void and candidate material, respectively. $\tilde{\rho_i}$ is the filtered density of the $\rho_i$, and $p$ is the SIMP parameter. Candidate materials with $E_1$ and $E_2$ Youngs' moduli (three-phase), Eq.~\ref{Eq:oneMaterial} can be written as
\begin{equation}~\label{Eq:twoMaterial}
	E_i = (1 -\tilde{\rho}_{i1}^p)E_\mathrm{min} + \tilde{\rho}_{i1}^p \left(\left(1-\tilde{\rho}_{i2}^p \right)E_1 + \tilde{\rho}_{i2}^p E_2\right)
\end{equation}
where $E_\text{min} = 10^{-6}\times\min(E_1,\,E_2)$. $\tilde{\rho}_{i1}$ decides the solid and void phases of the material layout. $\tilde{\rho}_{i1} =0$ ensures the void phase, whereas $\tilde{\rho}_{i1} =1$ gives the solid phase. $\tilde{\rho}_{i1} =1$ and $\tilde{\rho}_{i2} =1$ imply the second material, whereas $\tilde{\rho}_{i1} =1$ and $\tilde{\rho}_{i2} =0$ indicate the first material. $\tilde{\rho}_{i1}$ can be considered as the topology variable and $\tilde{\rho}_{i2}$ is the material selection variable. The total volume of the solid phase is determined as
\begin{equation}
	V_s = \displaystyle \sum_{i=1}^{{Nel}} \frac{v_i\tilde{\rho}_{i1}}{{Nel}}
\end{equation}
where $v_i$ is the volume of element~$i$ and ${Nel}$ is the total number of elements used to parameterize the design domain. From Eq.~\ref{Eq:twoMaterial}, the density of material 1 is $\tilde{\rho}_{i1}\left(1-\tilde{\rho}_{i2}\right)$, and that of material 2 is $\tilde{\rho}_{i1}\tilde{\rho}_{i2}$. The number of design variables is doubled in the three-phase material model (Eq.~\ref{Eq:twoMaterial}) compared to the two-phase material model (Eq.~\ref{Eq:oneMaterial}). 
Likewise, one can determine the interpolated Youngs' modulus for the four-phase conditions~\cite{sivapuram2021design}.

To circumvent the checkerboard patterns and apply the minimum length scale control on the members of the optimized designs, we employ the classical density filter \cite{bruns2001topology} herein. The filtered design variable $\tilde{\rho}_i$ of $\rho_i$ is defined as

\begin{equation}\label{Eq:densityfilter}
	\tilde{\rho_i} = \frac{\sum_{j=1}^{Nel} v_j \rho_j w(\mvect{x}_j)}{\sum_{j=1}^{Nel} v_j w(\mvect{x}_j)},
\end{equation}
where $v_j$ is the volume of neighboring element~$j$. $w(\mvect{x}_j)= \max\left(0,\,1-\frac{||\mvect{x}_i -\mvect{x}_j||}{r_\text{fill}}\right)$, is the weight function, wherein $||(.)||$ is a Euclidean  distance between centroids $\mvect{x}_i$ and $\mvect{x}_j$ of elements  $i$ and $j$, respectively. $r_\text{fill}$ is the filter radius. In matrix form, Eq.~\ref{Eq:densityfilter} is written as
\begin{equation}\label{Eq:densityfiltermatrix}
	\bm{\tilde{\rho}} = \mathbf{H}\bm{\rho}
\end{equation} 
The derivative of $\tilde{\rho_i}$ (Eq.~\ref{Eq:densityfilter}) with respect to $\rho_j$  can be evaluated as
\begin{equation}\label{Eq:derivativefilteractual}
	\pd{\tilde{\rho_i}}{\rho_j} = \frac{v_j w(\mvect{x}_j)}{\sum_{k=1}^{Nel}v_k w(\mvect{x}_k)},\, i.e.,\, \pd{\tilde{\bm{\rho}}}{\bm{\rho}} = \mathbf{H}^\top.
\end{equation}
Note that at the beginning of the optimization, we evaluate $\mathbf{H}$ (Eq.~\ref{Eq:densityfiltermatrix}) and store it as it does not change with the optimization iterations. 

\section{Pressure load modeling}\label{Sec:PressLoadModel}
This section presents the pressure modeling technique--the Darcy model with the conceptualized drainage term. Interested readers can refer to \cite{kumar2020topology,kumar2020topology3Dpressure} for a complete read for one material. The approach is extended herein for multi-material modeling. We know the pressure boundary conditions, i.e., a pressure difference across the domain. And also, in the initial stage of the optimization, elements can be treated as a porous medium. Therefore, a ubiquitous choice to model pressure field is the Darcy law~\cite{kumar2020topology}, wherein the flux $\bm{q}$ is determined  in terms of the permeability of medium $\kappa$, the fluid viscosity $\mu$, and the pressure difference $\nabla p$ as:
\begin{equation}\label{Eq:Darcyflux}
	\bm{q} = -\frac{\kappa}{\mu}\nabla p = -K(\tilde{\rho}) \nabla p,
\end{equation}
where $\tilde{\rho}$ is the filtered design variable (Eq.~\ref{Eq:densityfiltermatrix}). $K(\tilde{\rho})$, the elemental flow coefficient, is determined for element~$i$ as
\begin{equation}\label{Eq:Flowcoefficient}
	K(\tilde{\rho}_i) = K_v\left(1-(1-\epsilon) \mathcal{H}(\tilde{\rho}_{i1},\,\beta_\kappa,\,\eta_\kappa)\right),
\end{equation}
where
$\mathcal{H}(\tilde{{\rho}_{i1}},\,\beta_\kappa,\,\eta_\kappa) = \frac{\tanh{\left(\beta_\kappa\eta_\kappa\right)}+\tanh{\left(\beta_\kappa(\tilde{\rho}_{i1} - \eta_\kappa)\right)}}{\tanh{\left(\beta_\kappa \eta_\kappa\right)}+\tanh{\left(\beta_\kappa(1 - \eta_\kappa)\right)}}$. $K_s$ is the flow coefficient of the solid state, and that of the void phase is $K_v$. $\epsilon=\frac{K_s}{K_v}$ is called flow contrast \cite{kumar2020topology3Dpressure}. $\left\{ \eta_\kappa,\,\beta_\kappa\right\}$ are called the flow parameters, which define the step position and slope of $K(\tilde{\rho}_i)$ respectively. Remark that in the case of multi-material modeling, we note that $\tilde{{\rho}}_{i1}$ decides the solid and void state of element~$i$; therefore, the flow coefficient of element~$i$ (Eq~\ref{Eq:Flowcoefficient}) is defined only in terms of $\tilde{{\rho}}_{i1}$ and is independent of the number of candidate materials are taken for the optimization.

It is noted in~\cite{kumar2020topology,kumar2020topology3Dpressure,kumar2023TOPress} that the Darcy law alone may not provide the meaningful pressure drop for a TO setting; therefore, we introduce a drainage term, $Q_\text{drain}$, that helps achieve the suitable pressure field variation within the design domain as TO advances. $Q_\text{drain}$ is mathematically defined as
\begin{equation}
	{Q}_\text{drain} = -D(\tilde{\rho}_e) (p - p_{\text{ext}})
\end{equation}
The final balanced equation of the Darcy formulation with the drainage term can be written as:
\begin{equation}\label{Eq:stateequation}
	\nabla\cdot\bm{q} -Q_\text{drain} = 0.
\end{equation}
When Eq.~\ref{Eq:stateequation} is solved using the fundamentals of the finite element analysis and by applying the boundary conditions provided on the pressure load, we have
\begin{equation}\label{Eq:FinalbalanceEqu}
	\mvect{Ap} = \mvect{0},
\end{equation}
where $\mvect{A}$ and $\mvect{p}$ are the global flow matrix and pressure vector respectively. By solving Eq.~\ref{Eq:FinalbalanceEqu}, one gets the pressure field within the design domain as TO progresses. Now, this obtained field is converted to the consistent nodal loads as
\begin{equation}\label{Eq:nodalforce}
	\mvect{F} = -\mvect{T}\mvect{p},
\end{equation}
where $\mvect{F}$ is the global force vector, and $\mvect{T}$ is the transformation matrix \cite{kumar2020topology}. By using the modified Darcy law, we get the pressure field, i.e., an implicit description of the loading surface and consistent loads for the TO formulation, which is discussed next.

\section{Topology optimization formulation}\label{Sec:TOformulation}

As mentioned before, we use the density-based formulation~\cite{sigmund2013topology} for TO, wherein the design domain is described by hexagonal elements~\cite{kumar2022honeytop90}.  We consider three-phase pressure load problems. The number of design variables for each element is the same as that of the candidate materials employed, i.e., two-variable for each element herein. The following optimization problem  is solved with the volume constraints:

\begin{equation}\label{Eq:Optimizationequation}
	\begin{rcases}
		& \underset{\bm{\rho}}{\text{min}}
		& &{\mathbf{u}^\top\mathbf{Ku}}\\
		& \text{such that:}  &&\,\, \mathbf{Ap} = \mathbf{0 }\\
		&  &&\,\,\mathbf{Ku = F} = -\mathbf{T p}\\
		& \text{volume constraint:} && \,\,\displaystyle \sum_{i=1}^{\texttt{Nel}}v_i\tilde{\rho}_{i1}\le \left(v_{f_1} + v_{f_2}\right) \sum_{i=1}^{\texttt{Nel}}v_i\\
		&  && \displaystyle \sum_{i=1}^{\texttt{Nel}}v_i\tilde{\rho}_{i2}\le v_{f_2}\sum_{i=1}^{\texttt{Nel}}v_i\\
		&  && \mathbf{0}\le\bm{\tilde{\rho}}\le \mathbf{1}\\
		& \text{Data:} && v_{f_1},\,v_{f_2}, E_1,\, E_2, p
	\end{rcases},
\end{equation}
where $\mathbf{u}$ is the global displacement vector, and $\mathbf{K}$ is the global stiffness matrix which is evaluated by assembling the elemental stiffness matrix. As mentioned before, $\tilde{\rho}_{i1}$ decides the topology of the optimized design, whereas $\tilde{\rho}_{i2}$ determines material selection. Two linear volume constraints are formulated using $\tilde{\rho}_{i1}$ and $\tilde{\rho}_{i2}$.   The first volume constraint is used to control the total amount of the solid phase in the final topology, whereas the second is employed to fix only solid phase~2. Alternatively, one may also choose to use separate volume fractions for each phase; however, those constraints will be nonlinear in $\tilde{\rho}_{i1}$ and $\tilde{\rho}_{i2}$. 

The optimization problem noted in Eq~\ref{Eq:Optimizationequation} is solved using a gradient-based optimizer, the Method of Moving Asymptotes (MMA, cf.~\cite{svanberg1987method}). Therefore, sensitivities of the objective and constraints are needed with respect to the design variables. The sensitivities are determined using the adjoint-variable method in view of Sec.~\ref{Sec:MultimaterialModeling} and Sec.~\ref{Sec:PressLoadModel}. One can find a complete description of the sensitivities in \cite{kumar2020topology,kumar2020topology3Dpressure,kumar2023TOPress}.
\section{Numerical examples and discussions}\label{Sec:NumExDiss}

\begin{figure}
	\centering
	\begin{subfigure}[t]{0.45\textwidth}
		\centering
		\includegraphics[scale=0.8]{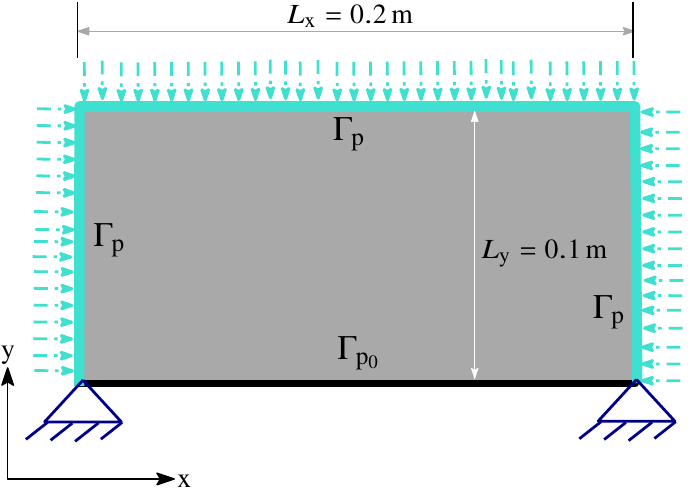}
		\caption{}
		\label{fig:lid}
	\end{subfigure}
	\quad
	\begin{subfigure}[t]{0.45\textwidth}
		\centering
		\includegraphics[scale=0.8]{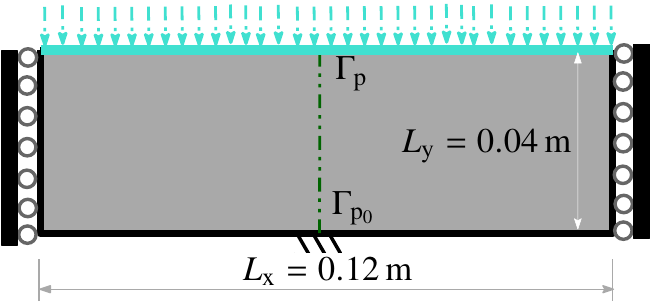}
		\caption{}
		\label{fig:piston}
	\end{subfigure}
	\caption{Design domains (\subref{fig:lid})  Externally pressurized  arch structure and (\subref{fig:piston}) Piston structure. Displacement and pressure boundary conditions are also depicted. $\Gamma_\text{p}$ and $\Gamma_\mathrm{p_0}$ indicate edges with $\SI{1}{\bar}$ and $\SI{0}{\bar}$ pressure loads, respectively.} \label{fig:DesigndomainsStrucutres}
\end{figure}

In this section, we present pressure-loaded design problems with two and three materials to demonstrate the success and efficacy of the presented method. We solve externally pressurized arch and pressure-loaded piston designs (Fig.~\ref{fig:DesigndomainsStrucutres}). One may also consider other benchmark problems from Ref.~\cite{kumar2023TOPress} to solve. The flow contrast $\epsilon = \SI{1e-7}{}$ is used. The out-of-plane thickness is fixed to $0.001 \si{\meter}$, and the plane-stress conditions are assumed. Poisson's ratio $\nu = 0.40$ is used. The flow parameters $\left\{\eta_\kappa,\,\beta_\kappa\right\} = \left\{0.20,\,10\right\}$ and the drainage parameters $\left\{\eta_d,\,\beta_d\right\} = \left\{0.20,\,10\right\}$ are taken.

Figure~\ref{fig:DesigndomainsStrucutres} displays the design domains of the externally pressurized arch and pressure-loaded piston structures. $\Gamma_\text{p}$ and $\Gamma_\mathrm{p_0}$ indicate the full and zero pressure load boundaries, respectively. The displacement boundary conditions are also depicted in the figure. $L_x = 0.2\si{\meter}$ and $L_y = 0.1\si{\meter}$ are taken for the arch design~(Fig.~\ref{fig:lid}), whereas for the piston design, $L_x = 0.12\si{\meter}$ and $L_y = 0.04\si{\meter}$ are set~(Fig.~\ref{fig:piston}). The arch design is solved considering the full model; this is performed to note any deviation from the symmetric nature of the result. However,  we exploit the symmetry nature of the piston design; thus, we analyze and optimize only its symmetric half part. $N_\text{ex}\times N_\text{ey} = 200 \times 100$ FEs and $N_\text{ex}\times N_\text{ey} = 180 \times 120$ FEs are used to describe the arch and piston domains, respectively. $N_\text{ex}$ and $N_\text{ey}$ are respectively the numbers of hexagonal elements in $x-$ and $y-$directions.  $r_\text{fill} = 3\times\left(\frac{L_x}{N_\text{ex}},\, \frac{L_y}{N_\text{ey}}\right)$ and $r_\text{fill} = 3.6 \sqrt 3\times\left(\frac{L_x}{N_\text{ex}},\, \frac{L_y}{N_\text{ey}}\right)$ are employed for the arch and piston structures. Different $r_\text{fill}$ are employed to achieve different minimum length scales for the problems. The maximum number of the MMA iterations is fixed to 100. The external move limit for the MMA optimizer is set to 0.1~\cite{svanberg1987method}.

\subsection{Two-material cases}
Here, we consider two candidate materials with $E_1 = \SI{40}{\mega \pascal}$ and  $E_2 = \SI{100}{\mega \pascal}$ for designing the arch and piston structures.  $v_{f_1} = 0.1$ and $v_{f_2}= 0.1$ are considered.

\begin{figure}
	\centering
	\begin{subfigure}[t]{0.45\textwidth}
		\centering
		\includegraphics[scale=0.5]{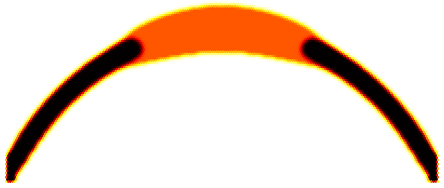}
		\caption{}
		\label{fig:lidmaterial}
	\end{subfigure}
	\quad
	\begin{subfigure}[t]{0.45\textwidth}
		\centering
		\includegraphics[scale=0.5]{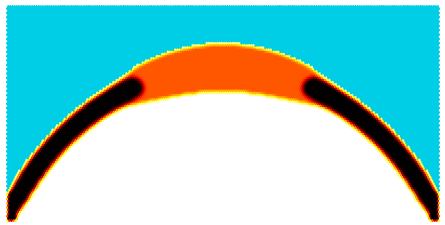}
		\caption{}
		\label{fig:lidpressmaterial}
	\end{subfigure}
	\caption{Optimized results of the externally pressurized arch structure with two candidate materials (\subref{fig:lidmaterial}) optimized material layout (\subref{fig:lidpressmaterial}) optimized material layout with the final pressure field. Black$\to$ Material~2~($E_2~=~\SI{100}{\mega \pascal}$ ),\, Orange$\to$ Material~1~($E_1 = \SI{40}{\mega \pascal}$).} \label{fig:lidresults}
\end{figure}
\begin{figure}
	\centering
	\begin{subfigure}[t]{0.45\textwidth}
		\centering
		\includegraphics[scale=0.5]{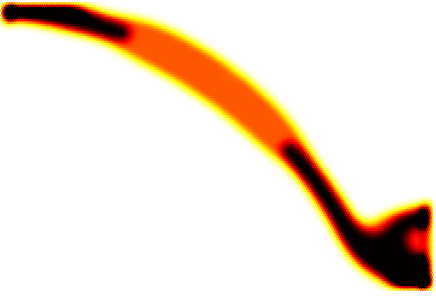}
		\caption{}
		\label{fig:pistonmaterial}
	\end{subfigure}
	\quad
	\begin{subfigure}[t]{0.45\textwidth}
		\centering
		\includegraphics[scale=0.5]{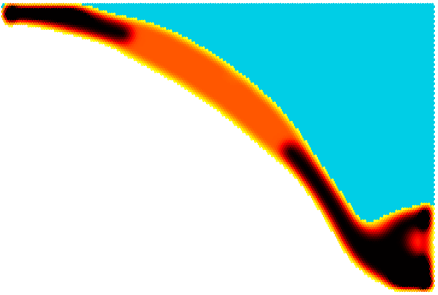}
		\caption{}
		\label{fig:pistonpressmaterial}
	\end{subfigure}
	\caption{Symmetric half Optimized pressure loadbearing piston design with two candidate materials (\subref{fig:pistonmaterial})  optimized material layout and (\subref{fig:pistonpressmaterial}) optimized material layout with the final pressure field. Black$\to$ Material~2~($E_2~=~\SI{100}{\mega \pascal}$ ),\, Orange$\to$ Material~1~($E_1 = \SI{40}{\mega \pascal}$).} \label{fig:pistonresults}
\end{figure}

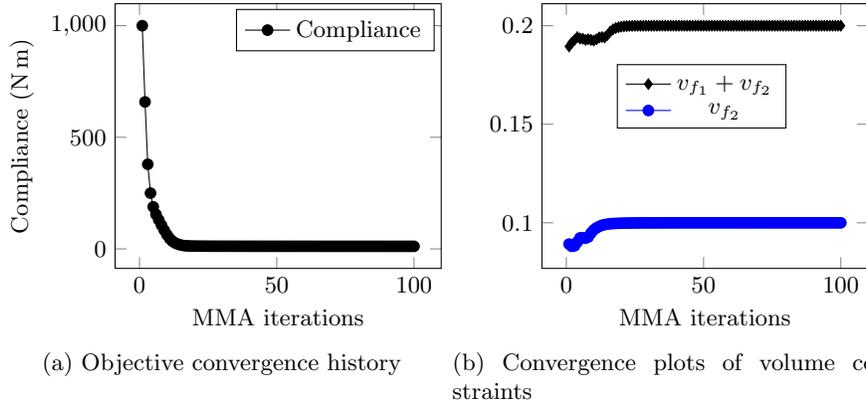
\begin{figure}
	\centering
	\begin{subfigure}[t]{0.485\textwidth}
		\centering
		\begin{tikzpicture} 	
			\pgfplotsset{compat = 1.3}
			\begin{axis}[
				width = 1\textwidth,
				xlabel=MMA iterations,
				ylabel= Compliance ($\si{\newton\meter}$)]
				\pgfplotstableread{objarch2MM.txt}\mydata;
				\addplot[mark=*,black]
				table {\mydata};
				\addlegendentry{Compliance}
			\end{axis}
		\end{tikzpicture}
		\caption{Objective convergence history}
		\label{fig:archobjhistory}
	\end{subfigure}
	\begin{subfigure}[t]{0.485\textwidth}
		\centering
		\begin{tikzpicture} 	
			\pgfplotsset{compat = 1.3}
			\begin{axis}[
				width = 1\textwidth,
				xlabel=MMA iterations,
				ylabel=  ,
				legend style={at={(0.75,0.65)},anchor=east}]
				\pgfplotstableread{vol1arch2MM.txt}\mydata;
				\addplot[mark=diamond*,black]
				table {\mydata};
				\addlegendentry{$v_{f_1}+v_{f_2}$}
				\pgfplotstableread{vol2arch2MM.txt}\mydata;
				\addplot[mark=otimes*,blue]
				table {\mydata};
				\addlegendentry{$v_{f_2}$ }
			\end{axis}
		\end{tikzpicture}
		\caption{Convergence plots of volume constraints}
		\label{fig:archvolumehistory}
	\end{subfigure}
	\caption{Convergence plots of the objective and volume constraints of the arch problem.}\label{fig:archobjvolumplots}
\end{figure}

Fig.~\ref{fig:lidresults} and Fig.~\ref{fig:pistonresults} depict optimized loadbearing arch and piston results. The optimized designs contain both materials (three-phase: void, material~1, and material~2). In addition, the symmetric nature of the arch structure is maintained in its optimized design (Fig.~\ref{fig:lidresults}). The optimizer places material with high elastic stiffness at the displacement boundary to reduce the deformation, and hence, it minimizes the objective. Fig.~\ref{fig:lidpressmaterial} and Fig.~\ref{fig:pistonpressmaterial} depict the optimized material layouts of the arch and piston structures with the final pressure fields. The optimizer provides the final material layouts in such a manner that they contain the pressure loads as well as form the stiffest designs. As in TO approaches, the boundaries of the optimized designs are constituted via the edges of the FEs employed to describe the domains. Therefore, typically, optimized results contain undulated boundaries. To achieve relatively smooth boundaries of the optimized designs, one can use the boundary smoothing scheme per~\cite{kumar2015topology} within the proposed approach. The two-material arch design's objective and constraints history plots are shown in Fig.~\ref{fig:archobjvolumplots}. The objective convergence for the problem is rapid and smooth. One can note that both volume constraints remain active at the end of optimization.
\subsection{Three-material cases}

For three candidate materials, we consider $E_1 = \SI{10}{\mega \pascal}$, $E_2 = \SI{40}{\mega \pascal}$ and  $E_3 = \SI{100}{\mega \pascal}$ for designing pressure loadbearing arch and piston structures.  $v_{f_1} =0.1 $, $v_{f_2}= 0.1$ and $v_{f_3}= 0.05$ are set.
\begin{figure}
	\centering
	\begin{subfigure}[t]{0.45\textwidth}
		\centering
		\includegraphics[scale=0.5]{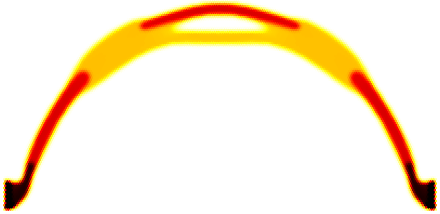}
		\caption{}
		\label{fig:lidmaterial3}
	\end{subfigure}
	\quad
	\begin{subfigure}[t]{0.45\textwidth}
		\centering
		\includegraphics[scale=0.5]{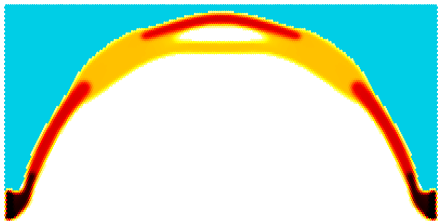}
		\caption{}
		\label{fig:lidpressmaterial3}
	\end{subfigure}
	\caption{Optimized results of the externally pressurized arch structure with three candidate materials (\subref{fig:lidmaterial}) optimized material layout (\subref{fig:lidpressmaterial}) optimized material layout with the final pressure field. Black$\to$ Material~3~($E_3~=~\SI{100}{\mega \pascal}$ ),\, Orange$\to$ Material~2~($E_2 = \SI{40}{\mega \pascal}$),\,Gold$\to$ Material~1~($E_1 = \SI{10}{\mega \pascal}$)} \label{fig:lidresults3}
\end{figure}

\begin{figure}
	\centering
	\begin{subfigure}[t]{0.45\textwidth}
		\centering
		\includegraphics[scale=0.5]{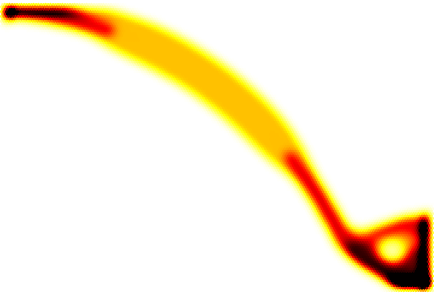}
		\caption{}
		\label{fig:pistonmaterial3}
	\end{subfigure}
	\quad
	\begin{subfigure}[t]{0.45\textwidth}
		\centering
		\includegraphics[scale=0.5]{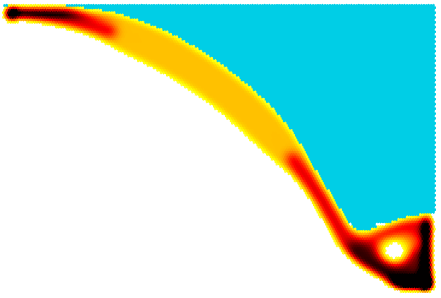}
		\caption{}
		\label{fig:pistonpressmaterial3}
	\end{subfigure}
	\caption{Symmetric half optimized pressure loadbearing piston design with three candidate materials (\subref{fig:pistonmaterial})  optimized material layout and (\subref{fig:pistonpressmaterial}) optimized material layout with the final pressure field. Black$\to$ Material~3~($E_3~=~\SI{100}{\mega \pascal}$ ),\, Orange$\to$ Material~2~($E_2 = \SI{40}{\mega \pascal}$),\,Gold$\to$ Material~1~($E_1 = \SI{10}{\mega \pascal}$).} \label{fig:pistonresults3}
\end{figure}

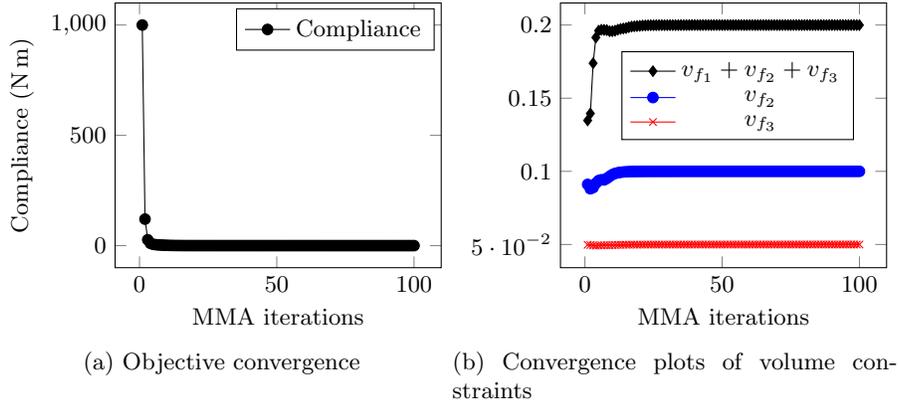
\begin{figure}
	\centering
	\begin{subfigure}[t]{0.485\textwidth}
		\centering
		\begin{tikzpicture} 	
			\pgfplotsset{compat = 1.3}
			\begin{axis}[
				width = 1\textwidth,
				xlabel=MMA iterations,
				ylabel= Compliance ($\si{\newton\meter}$)]
				\pgfplotstableread{objpiston3MM.txt}\mydata;
				\addplot[mark=*,black]
				table {\mydata};
				\addlegendentry{Compliance}
			\end{axis}
		\end{tikzpicture}
		\caption{Objective convergence}
		\label{fig:pistonobjhistory}
	\end{subfigure}
	\begin{subfigure}[t]{0.485\textwidth}
		\centering
		\begin{tikzpicture} 	
			\pgfplotsset{compat = 1.3}
			\begin{axis}[
				width = 1\textwidth,
				xlabel=MMA iterations,
				ylabel=  ,
				legend style={at={(0.90,0.65)},anchor=east}]
				\pgfplotstableread{vol1piston3MM.txt}\mydata;
				\addplot[mark=diamond*,black]
				table {\mydata};
				\addlegendentry{$v_{f_1} + v_{f_2} + v_{f_3}$}
				\pgfplotstableread{vol2piston3MM.txt}\mydata;
				\addplot[mark=otimes*,blue]
				table {\mydata};
				\addlegendentry{$v_{f_2}$ }
				\pgfplotstableread{vol3piston3MM.txt}\mydata;
				\addplot[mark=x,red]
				table {\mydata};
				\addlegendentry{$v_{f_3}$ }
			\end{axis}
		\end{tikzpicture}
		\caption{Convergence plots of  volume constraints}
		\label{fig:pistonvolumehistory}
	\end{subfigure}
	\caption{Convergence plots of the objective and volume constraints of the three-material piston problem.}\label{fig:pistonobjvolumplots}
\end{figure}

The optimized loadbearing arch and piston designs with three-material are depicted in Fig.~\ref{fig:lidmaterial3} and Fig.~\ref{fig:pistonmaterial3}, respectively. The final pressure fields with the optimized material layout for these designs are also shown in Fig.~\ref{fig:lidpressmaterial3} and Fig.~\ref{fig:pistonpressmaterial3}, respectively. The optimized shape of the arch design maintains symmetry. Material with $E_3=\SI{100}{\mega \pascal}$ (black) forms the fixed boundaries in the optimized designs. The convergences for objective and volume constraints for the piston problem are depicted in Fig.~\ref{fig:pistonobjvolumplots}. The objective convergence is smooth and rapid, as observed for the two-material arch problem (Fig.~\ref{fig:archobjvolumplots}). The volume constraints remain active at the end of the optimization.
\section{Concluding remarks}\label{Sec:ConRem}
This paper presents a topology optimization approach to designing fluidic pressure-loaded structures with multi-material. The versatility of the proposed method is demonstrated by solving two structural problems subjected to fluidic pressure loads with two and three candidate designs. The extended SIMP approach is employed for the modeling of the multi-material. Compliance is minimized with two and three linear volume constraints for two and three candidate materials. Honeycomb tessellation is used to describe the design domain that inherently provides nonsingular connectivity between two juxtaposed elements. In the optimized designs, the symmetric nature of the problems remains intact. In the optimized designs, the material with the highest elastic stiffness forms the fixed boundary conditions, thus reducing the deformation and compliance of the structures. The objective convergence is smooth and rapid. Volume constraints remain active at the end of the optimization.

Pressure load is modeled using the Darcy law with the conceptualized drainage term as per~\cite{kumar2020topology}, wherein the flow coefficient of an element is determined using its solid and void phases irrespective of the number of candidate materials. Likewise, the drainage term is defined. The presented scheme works well, as noted by solving two problems. In the near future, we envision to present a topology optimization approach for pressure-driven multi-material compliant mechanisms.

\end{document}